\documentclass[a4paper, USenglish, cleveref, thm-restate]{lipics-v2021}

\title{A Simple Algorithm for Combinatorial n-Fold ILPs Using the Steinitz Lemma} 
\nolinenumbers

\author{Sushmita Gupta}{The Institute of Mathematical Sciences, HBNI, Chennai, India}{sushmitagupta@imsc.res.in}{https://orcid.org/0000-0003-1255-8266}{}
\author{Pallavi Jain}{IIT Jodhpur, India}{pallavi@iitj.ac.in}{https://orcid.org/0000-0001-8900-9797}{}
\author{Sanjay Seetharaman}{The Institute of Mathematical Sciences, HBNI, Chennai, India}{sanjays@imsc.res.in}{https://orcid.org/0009-0001-1483-6138}{}
\author{Meirav Zehavi}{Ben-Gurion University of the Negev, Beer-Sheva, Israel}{meiravze@bgu.ac.il}{https://orcid.org/0000-0002-3636-5322}{}

\authorrunning{S. Gupta, P. Jain, S. Seetharaman, and M. Zehavi} 

\Copyright{Sushmita Gupta, Pallavi Jain, Sanjay Seetharaman, and Meirav Zehavi} 

\ccsdesc[100]{Theory of computation~Integer programming}

\keywords{$n$-fold integer linear program, parameterized algorithms} 

\acknowledgements{We thank the reviewers of an earlier draft as well as this for their contribution in improving the clarity and presentation of our results.}

\EventEditors{Akanksha Agrawal and Erik Jan van Leeuwen}
\EventNoEds{2}
\EventLongTitle{20th International Symposium on Parameterized and Exact Computation (IPEC 2025)}
\EventShortTitle{IPEC 2025}
\EventAcronym{IPEC}
\EventYear{2025}
\EventDate{September 17--19, 2025}
\EventLocation{Warsaw, Poland}
\EventLogo{}
\SeriesVolume{358}
\ArticleNo{14}

\usepackage{algorithm}
\usepackage{algpseudocode}
\usepackage{thmtools}
\usepackage{mathtools}
\usepackage{todonotes}
\usepackage{xspace}
\usepackage{tabularx}
\newcommand{\newdef}[1]{{\bf #1}}
\newcommand{\etal}{et al.\xspace}

\DeclareMathSymbol{\qm}{\mathalpha}{operators}{"3F}
\DeclareMathAlphabet{\mathbbold}{U}{bbold}{m}{n}

\newcommand\bigoh{\mathcal{O}}

\newcommand{\A}{\ensuremath{\mathscr{A}}}
\newcommand{\LP}{\ensuremath{\mathbf{LP}}\xspace}

\newcommand{\imb}{\textup{\texttt{imb}}}
\newcommand{\occ}{\textup{\texttt{occ}}}

\newcommand{\isize}[1]{\ensuremath{\langle #1\rangle}}

\newcommand{\closeststring}{\textsc{Closest String}\xspace}

\newcommand{\dms}{\textsc{$\delta$-Multi Strings}\xspace}

\newcommand{\nfILP}{$n$-fold ILP\xspace}

\newcommand{\ma}[1]{\todo[color=green!70!blue!30!]{\small{#1}}}

\newcommand{\Ma}[1]{\textcolor{magenta}{#1}}

\newcommand{\defproblem}[3]{
  \vspace{1mm}
  \noindent\fbox{
    \begin{minipage}{0.95508\textwidth}
      \begin{tabular*}{\textwidth}{@{\extracolsep{\fill}}lr} #1 \\ \end{tabular*}
      {\bf{Input:}} #2  \\
      {\bf{Question:}} #3
    \end{minipage}
  }
  \vspace{1mm}
}

\newcommand{\FPT}{{\sf FPT}\xspace}

\newcommand{\NPc}{{\sf NP}-{\sf complete}\xspace}
\newcommand{\NP}{{\sf NP}\xspace}

\newcommand{\hide}[1]{{}}

\newcommand{\trans}{\intercal}
\newcommand{\psum}{\mathrm{psum}}
\newcommand{\tot}{\mathrm{tot}}

\newcommand{\equitablecoloring}{\textsc{Equitable Coloring}\xspace}
\newcommand{\lobbying}{\textsc{Lobbying}\xspace}

\declaretheorem[numberwithin=theorem,name=Claim]{claim-inside-theorem}
\Crefname{claim-inside-theorem}{Claim}{Claims}
\declaretheorem[numberlike=claim-inside-theorem,name=Lemma]{lemma-inside-theorem}
\Crefname{lemma-inside-theorem}{Lemma}{Lemmas}
\declaretheorem[numberlike=claim-inside-theorem,name=Observation]{observation-inside-theorem}
\declaretheorem[numberwithin=lemma,name=Claim]{claim-inside-lemma}
\Crefname{claim-inside-lemma}{Claim}{Claims}
\declaretheorem[numberlike=claim-inside-lemma,name=Corollary]{corollary-inside-lemma}
\declaretheorem[numberwithin=lemma,name=Lemma]{lemma-inside-lemma}
\declaretheorem[numberlike=claim-inside-lemma,name=Observation]{observation-inside-lemma}
\declaretheorem[numberwithin=lemma-inside-theorem,name=Claim]{claim-inside-lemma-inside-theorem}
\declaretheorem[numberwithin=lemma-inside-theorem,name=Lemma]{lemma-inside-lemma-inside-theorem}
\declaretheorem[numberwithin=lemma-inside-theorem,name=Observation]{observation-inside-lemma-inside-theorem}

\declaretheorem[name=Construction,style=remark,shaded={rulewidth=0pt}]{construction}

\begin{document}

\maketitle
\allowdisplaybreaks


\begin{abstract}
We present an algorithm for a class of $n$-fold ILPs whose existing algorithms in literature are often either (1) based on the \textit{augmentation framework} where one starts with an arbitrary solution and then iteratively moves towards an optimal solution by solving appropriate programs; or (2) require solving a linear relaxation of the program; or (3) are based on decomposition/proximity based arguments.
Combinatorial $n$-fold ILPs is a class of $n$-fold ILPs introduced and studied by Knop et al. [MP2020] that captures several other problems in a variety of domains.
We present a simple and direct algorithm that solves combinatorial $n$-fold ILPs with unbounded non-negative variables via an application of the Steinitz lemma.
Depending on the structure of the input ILP, we also improve upon the existing algorithms in the literature in terms of the running time, thereby showing an improvement that mirrors the one shown by Rohwedder [ICALP2025] contemporaneously and independently.
\end{abstract}

\section{Introduction}
The power of the {\it linear
algebra method} has a long and storied history in the design and analysis of algorithms, \cite{Bárány2008,DBLP:journals/dam/Sevastjanov94}. 
The question of reordering vectors so that all partial sums are bounded was posed by mathematicians Paul L\'evy and George Riemann in the early 20th century. \hide{Specifically, given a finite collection of vectors in $\mathbb{R}^d$ that \Ma{sum to zero and are bounded in norm}, they asked whether there always exists a reordering such that all partial sums lie within a fixed bound.\ma{is this the exact version that was asked by L\&R?}} 
The question was first affirmatively answered by Ernst Steinitz \cite{Steinitz,Grinberg1980} showing that such a reordering of vectors, 
that sum to zero and are all in a unit ball with respect to any arbitrary norm, is indeed possible where all the partial sums can be bounded by a constant that depends only on $d$, the dimension of the vectors. 
This result, now eponymously known as the Steinitz Lemma, stated in \Cref{thm:steinitz-lemma}, has become a cornerstone in our understanding of the geometry of numbers and has led to numerous algorithmic applications centered around problems that have an integer linear program (ILP) formulation \cite{DBLP:journals/talg/EisenbrandW20}. 
In this paper, we study a simple Steinitz Lemma based non-augmenting algorithm for a class of combinatorial $n$-fold ILPs. 

The usefulness of Steinitz Lemma in modern algorithmic research is perhaps best underscored by its applicability to analyze integer linear programs (ILP), 
as exhibited by Eisenbrand and Weismantel \cite{DBLP:journals/talg/EisenbrandW20} who improved upon the longstanding best bound of Papadimitriou~\cite{DBLP:journals/jacm/Papadimitriou81}\hide{\footnote{To the best of our knowledge, the arXiv and subsequent SODA'18 version of this paper was the first to use Steinitz Lemma to demonstrate such an improvement.
}}. 
Papadimitriou's seminal work gave a simpler proof of ILPs being in \NP and the first pseudo-polynomial algorithm when the number of constraints is constant. This can be viewed as a {\it pseudo-parameterized algorithm}\hide{--one whose time complexity may depend on size of the coefficients--} with respect to the number of constraints. 
Subsequently, Jansen and Rohwedder \cite{DBLP:conf/innovations/JansenR19} further improved the running time by applying the Steinitz Lemma in an alternative way.
In this paper, we are continuing this line of investigation by employing Steinitz Lemma on a special class of ILPs,  namely combinatorial \nfILP, for the purpose of obtaining better bounds.




Notwithstanding these exciting possibilities, the bottom line is that solving an ILP is known to be \NPc and that poses a computational challenge when we desire integral solutions only. 
While we know of certain classes of ILPs, such as {\it totally unimodular}, {\it bimodular} \cite{DBLP:conf/stoc/ArtmannWZ17}, {\it binet} \cite{DBLP:journals/orl/AppaKPP07} that can be solved in polynomial time, the task of mapping the ILP landscape into families that are tractable under modest restrictions is ongoing. In particular, in recent times researchers are motivated to design parameterized algorithms because they behave like polynomial-time algorithms when the parameter is constant-valued. 
An algorithm is said to be {\it fixed-parameter tractable} (\FPT) with respect to parameters $k_1, \dots, k_\ell$, if its running time is of the form $f(k_1, \dots, k_\ell)\cdot |\text{input}|^{\bigoh(1)}$ for some computable function $f$.
Steinitz Lemma has proven to be useful even in this direction of research, as exhibited in its usefulness in the study of {\it block-structured} ILPs. 

\subparagraph*{Block-structured ILP.} A class of block structured programs called the $n$-{\it fold} ILP has come to sharp focus due to their relevance to efficiently solving problems in scheduling~\cite{DBLP:journals/scheduling/KnopK18}, fair division~\cite{DBLP:conf/ijcai/0004R23}, and social choice~\cite{DBLP:journals/teco/KnopKM20} to name a few. 
For example, in the \textit{high-multiplicity} scheduling setting, the input is succinctly encoded by partitioning the jobs (machines) into \textit{types} such that all jobs (machines) within a type are identical.
Note that the number of job (machine) types can be much smaller than the number of jobs (machines).
Using \nfILP, Knop and Kouteck{\'{y}} \cite{DBLP:journals/scheduling/KnopK18} obtain algorithms for some scheduling problems that are FPT with respect to the number of job (machine) types.





More broadly, the $n$-fold setting has unlocked the door towards parameterized and approximation algorithms and their rich tool-kit. 
Formally, we define \nfILP as the following 
\[\min\left\{c^\trans x\, \mid \mathscr{A} x= b\,,\ell \leq x \leq u\,, x\in\mathbb{Z}^{nt}\right\}, \tag{$\mathbf{P_0}$} \label{prob:nfILP-with-bounds}\]
where vectors $c \in \mathbb{Z}^{nt}, b \in \mathbb{Z}^{r+ns}, \ell,u \in (\mathbb{Z} \cup \{-\infty, \infty\})^{nt}$ and the constraint matrix
 \begin{align*}
    \mathscr{A}:=
\left(
\begin{array}{cccc}
T    & T    & \cdots & T    \\
D    & 0      & \cdots & 0      \\
0      & D    & \cdots & 0      \\
\vdots & \vdots & \ddots & \vdots \\
0      & 0      & \cdots & D    \\
\end{array}
\right), \text{such that}
\end{align*}
the top block $T \in \mathbb{Z}^{r \times t}$ and the diagonal block $D \in \mathbb{Z}^{s \times t}$. Let $\Delta_X$ denote an upperbound on the absolute value of an entry in a matrix $X$, and let $\Delta_I$ denote the same for the numbers in the input.
Hemmecke~\etal\cite{DBLP:journals/mp/HemmeckeOR13} were the first to give an \FPT algorithm for \eqref{prob:nfILP-with-bounds} with respect to the parameters $\Delta_{\A},r,s,t$. This algorithm runs in time $\bigoh(n^3t^3 \cdot\Delta_{\A}^{\bigoh(t(rs+st))} \log(\Delta_{c})) $.
Hence, this has better performance than \cite{DBLP:journals/jacm/Papadimitriou81} and \cite{DBLP:journals/talg/EisenbrandW20} that work for general ILPs. 
Subsequently, Eisenbrand \etal\cite{DBLP:conf/icalp/EisenbrandHK18}, using the Steinitz Lemma in a manner similar to \cite{DBLP:journals/talg/EisenbrandW20}, improved the result of \cite{DBLP:journals/mp/HemmeckeOR13}, in terms of the dependence on $t$ given that $t\geq r,s$. Their algorithm has running time $n^2t^2\log\Delta_{I}\log^2{nt} \cdot (rs\Delta_{\A})^{\bigoh(r^2s+rs^2)} + \LP$, where \LP denotes the time needed to solve the LP relaxation of \eqref{prob:nfILP-with-bounds}. 
This is an \FPT algorithm with respect to the parameters $r,s,\Delta_\A$. 
We are naturally interested in algorithms with smaller dependence on $t$, the number of columns, because $t$ can be as large as $\Delta_{\A}^{r+s}$, a common occurrence in applications involving configuration IPs, as noted in \cite{DBLP:conf/icalp/EisenbrandHK18}. 
Koutecky et al. \cite{DBLP:conf/icalp/KouteckyLO18} showed an algorithm that runs in time $(nt)^6 \log (nt) \cdot (rs\Delta_\A )^{\bigoh(r^2s+rs^2)} + \LP$.
This is the first strongly  polynomial time algorithm for fixed values of $r,s, \Delta_\A$.
Subsequently, Eisenbrand et al. \cite{DBLP:journals/corr/abs-1904-01361}, Jansen et al. \cite{DBLP:journals/siamdm/JansenLR20}, and Cslovjecsek et al. \cite{DBLP:conf/soda/CslovjecsekEHRW21} have developed algorithms that run in time $(nt)^2 \log^3(nt) (rs\Delta_\A)^{\bigoh(r^2s+rs^2)}+\LP$, $nt \log^{\bigoh(1)} (nt) \cdot \log (\Delta_I^2) \cdot (rs\Delta_\A)^{\bigoh(r^2s+s^2)}$, and $(nt)^{1+o(1)} \cdot 2^{\bigoh(rs^2)} (rs\Delta_\mathscr{A})^{\bigoh(r^2s+s^2)}$, respectively.



A special class of \nfILP known as the {\it combinatorial $n$-fold ILP}--where the diagonal block $D=(1,\ldots,1) \in \mathbb{Z}^{1\times t}$--was introduced by Knop et al. \cite{DBLP:journals/mp/KnopKM20} and it captures many problems in various domains including computational social choice, stringology, and scheduling.
They showed that such programs can be solved in time $t^{\bigoh(r)}(\Delta_\A r)^{\bigoh(r^2)} \bigoh(n^3\cdot\isize{I})+\LP$, where $\isize{I}$ denotes the size of encoding $\langle b,\ell,u,c \rangle$. 
We would like to note that the result of \cite{DBLP:conf/soda/CslovjecsekEHRW21} improves upon this and gives a $(r\Delta_\A)^{\bigoh(r^2)} (nt)^{1+o(1)}$ algorithm for this special class.

In this paper, we solve a variant of combinatorial \nfILP where the variables are non-negative and unbounded from above. 
However, our constraint matrix is more general because in our setting the top blocks $T$ need not be identical. We formally define our problem and result in \Cref{ss:contribution}. 
We believe our methodology is conceptually simpler than the earlier approaches, described in the next sub-section, for solving this problem. 

\subsection{Background and our contribution}\label{ss:contribution}
In this article we present a simple non-iterative algorithm that leverages the power of the Steinitz Lemma to solve a class of combinatorial \nfILP, \eqref{eqn:prob-def}. 
In other words, our approach solves this problem directly and not via a sequence of \textit{augmentation} steps involving the \textit{Graver bases} as exhibited in earlier papers such as \cite{DBLP:journals/talg/EisenbrandW20,DBLP:journals/corr/abs-1904-01361}, or via non-iterative approaches that involve proximity-based arguments \cite{DBLP:conf/soda/CslovjecsekEHRW21} or  decomposition-based arguments \cite{DBLP:conf/soda/CslovjecsekKLPP24}. 


\hide{Our work in this paper, the performance and approach, sits here in the ladder of successively improved results for solving ILP based problems.}




Formally, we define our setting as follows. 

\subparagraph*{Our setting.}
We study ILPs of the following form
\begin{align*}
\label{eqn:prob-def}
\min\left\{c^\trans x\, \mid \mathscr{A} x= b\,, x\in\mathbb{Z}_{\ge 0}^{nt}\right\}\hspace{1.5cm}\tag{$\mathbf{P_1}$}\\
~~ \mbox{where vectors } b \in \mathbb{Z}^{r+n} \text{ and } c \in \mathbb{Z}^{nt \times 1}; \text{ and the constraint matrix}\\
\mathscr{A}:=
\left(
\begin{array}{cccc}
T^{(1)}    & T^{(2)}    & \cdots & T^{(n)}    \\
D    & 0      & \cdots & 0      \\
0      & D   & \cdots & 0      \\
\vdots & \vdots & \ddots & \vdots \\
0      & 0      & \cdots & D   \\
\end{array}
\right) \text{ such that } \notag
\end{align*}
the {\it diagonal} block $D = (1, \dots, 1) \in \mathbb{Z}^{1 \times t}$ and the {\it top} blocks  $T^{(1)}, \dots, T^{(n)} \in \mathbb{Z}^{r \times t}$. 

The main technical contribution of this paper is to prove the following.

\begin{theorem}\label{thm:main}
    The {\rm ILP} \eqref{eqn:prob-def} can be solved in time $\bigoh(nt \cdot (n\Delta_{\A}(n+1+4r))^r \cdot q )$, where 
\hide{$\Delta_{\A}$ denotes the largest absolute value of an entry in \A and} $q$ denotes the sum of the last $n$ entries in $b$. 
    
\end{theorem}





While the expression of time complexity in \Cref{thm:main} is not \FPT with respect to $r,\Delta_\A$, we note, however, that it allows us to identify a class of instances where our algorithm will outperform the existing algorithms  for \nfILP, including \cite{DBLP:journals/corr/abs-1904-01361,DBLP:journals/siamdm/JansenLR20,DBLP:conf/soda/CslovjecsekEHRW21}, as exhibited for the problems {\sc Lobbying} and \textsc{Binary} \closeststring in \Cref{sec:Applications}.
We consider this to be a conceptual contribution of our approach and note that structure of the $b$ vector, and parameters such as $q$, are worthy of further investigation.
We also present an application of our result on the problem \equitablecoloring.

We note that if we apply the {\it pre-n fold ILP} framework \cite[Corollary~23]{DBLP:journals/mp/KnopKM20} to \eqref{eqn:prob-def} we would obtain an algorithm with running time $(tn)^{r}(\Delta_{\A}r)^{\bigoh(r^2)}\bigoh(n^3 \cdot \isize{I}))+ \LP$.
There is no linear dependence on $q$, and there is an $(\Delta_{\A}r)^{\bigoh(r^2)}$ term in the running time.

Next, we state the running time of existing algorithms when applied to \eqref{eqn:prob-def}:
Eisenbrand et al. \cite{DBLP:conf/icalp/EisenbrandHK18} $n^2t^2 \log^{2} nt \log\Delta_{I}\cdot(r\Delta_{\A})^{\bigoh(r^2)} + \LP$;
Jansen et al. \cite{DBLP:journals/siamdm/JansenLR20} $nt \log^{\bigoh(1)}(nt) \cdot \log^2\Delta_I \cdot (r\Delta_\A)^{\bigoh(r^2)}$;
Cslovjecsek et al. \cite{DBLP:conf/soda/CslovjecsekEHRW21} $(nt)^{1+o(1)} 2^{\bigoh(r)} (r\Delta_\A)^{\bigoh(r^2)}$.
The algorithms in \cite{DBLP:journals/mp/HemmeckeOR13,DBLP:conf/icalp/KouteckyLO18,DBLP:conf/soda/CslovjecsekKLPP24} require the top row block to contain identical matrices and it is not clear how to obtain a solution to our problem using them.

\subparagraph*{Contemporaneous related works.}
In \cite{DBLP:conf/icalp/Rohwedder25} (recently accepted in ICALP'25), Rohwedder presented an algorithm for \textsc{Multi-choice Integer Programming}, a problem very closely related to~\eqref{eqn:prob-def} (either can be reduced to the other in polynomial time). 
The running times are identical (in terms of the dependence on parameters), and the techniques are very similar. 
The main technical difference is in the first phase of the algorithm (\Cref{lem:imbalance-bound-one-row}) where they present a ``time-based'' approach instead of the ``greedy imbalance-based'' approach we present.
The main result of that paper is an algorithm for scheduling on uniform machines, which uses the solution for the above problem as a subroutine. 
Our main contribution is a more detailed presentation of the algorithm, including in~\Cref{ss:dealing-with-inequalities} how we deal with inequalities in the constraints and apply it on \lobbying and \textsc{Binary Closest String}.
Moreover, our result can be applied to obtain the corollaries and applications indicated by Rohwedder~\cite{DBLP:conf/icalp/Rohwedder25}. We note that dealing with inequalities is an important step in our analysis because we are using the $n$-fold ILP framework while Rohwedder is not. 

We also note that Jansen et al. \cite{DBLP:journals/corr/abs-2409-04212} present a divide-and-conquer-based algorithm (based on the approach in \cite{DBLP:journals/mor/JansenR23}) for a problem that is closely related to the feasibility version of \eqref{eqn:prob-def}. 
It is not clear whether similar guarantees as ours could be achieved by their result, we refer to \cite{DBLP:conf/icalp/Rohwedder25} for a discussion on this.
We note that in the updated version \cite{jansen2025newalgorithmcombinatorialnfolds} of the work (also accepted in IPEC'25), they present an algorithm for \eqref{eqn:prob-def} with running time that has an identical exponential dependence on $n, r, \Delta_\A$, but has only a logarithmic dependence on $q$.
These articles are contemporaneous and independent.

\paragraph*{Preliminaries} 

The main technical tool in our result is the Steinitz Lemma, stated below.
\begin{proposition}[Steinitz Lemma,~\cite{Steinitz,Grinberg1980}]
    \label{thm:steinitz-lemma}
    Let $\lVert \cdot \rVert$ be an arbitrary norm of $\mathbb{R}^d$.
    Let $x_1, \dots, x_m \in \mathbb{R}^d$ such that $\sum_{i \in [m]} x_i = 0$ and $\lVert x_i \rVert \le 1$ for each $i \in [m]$.
    There exists a permutation $\pi \in S_m$ such that all partial sums satisfy
    \[
    \lVert \sum_{j \in [k]} x_{\pi(j)} \rVert \le d \text{ for each } k \in [m].
    \]
\end{proposition}

Eisenbrand and Weismantel \cite{DBLP:journals/talg/EisenbrandW20} generalize the above lemma and show the following.
\begin{proposition}[\cite{DBLP:journals/talg/EisenbrandW20}]
    \label{cor:steinitz-lemma-corollary}
    Let $x_1, \dots, x_m \in \mathbb{R}^d$ such that $\sum_{i \in [m]} x_i = s$ and $\lVert x_i \rVert_\infty \le \Delta$ for each $i \in [m]$.
    There exists a permutation $\pi \in S_m$ such that all partial sums satisfy
    \[
    \lVert \sum_{j \in [k]} x_{\pi(j)} - \frac{k}{m} \cdot s \rVert_\infty \le 2 \Delta d \text{ for each } k \in [m].
    \]
\end{proposition}





\hide{
\ma{Moved to our contribution
}
Let $A = (1, \dots, 1) \in \mathbb{Z}^{1 \times t}$, $T^{(1)}, \dots, T^{(n)} \in \mathbb{Z}^{r \times t}$, $b \in \mathbb{Z}^{r+n}$, and $c \in \mathbb{Z}^{nt \times 1}$. 
We study integer programs of the form
\begin{align}
\label{eqn:prob-def}
\min\left\{c^\trans x\, \mid \mathscr{A} x= b\,, x\in\mathbb{Z}_{\ge 0}^{nt}\right\}\tag{$\mathbf{P_1}$}\\
~~ \mbox{where }~\mathscr{A}:=
\left(
\begin{array}{cccc}
T^{(1)}    & T^{(2)}    & \cdots & T^{(n)}    \\
D    & 0      & \cdots & 0      \\
0      & D    & \cdots & 0      \\
\vdots & \vdots & \ddots & \vdots \\
0      & 0      & \cdots & D    \\
\end{array}
\right).\notag
\end{align}
**********************
}

\subparagraph*{Note on variations of \nfILP.} \eqref{eqn:prob-def} forms a special case of the $n$-fold integer programs. 
The following are some of the studied variations of such programs: 
\begin{enumerate}
\item The $n$ diagonal blocks, 
 which are $D=(1, \dots, 1)$ above, 
can be non-identical $s\times t$ matrices, \cite{DBLP:conf/icalp/EisenbrandHK18,DBLP:journals/mp/KnopKLMO23};

\item Moreover, {\it combinatorial} $n$-fold IPs usually refers to problems where the constraint matrix is a restricted version of \eqref{eqn:prob-def}, with the top block, denoted by $(T^{(1)}, \ldots, T^{(n)})$, being $n$ identical matrices,~\cite{DBLP:journals/mp/KnopKM20,DBLP:conf/soda/CslovjecsekKLPP24};


\item The objective function may be more general than linear, such as separable convex,~\cite{DBLP:journals/mp/KnopKM20}.
\end{enumerate}








\subparagraph*{Notation.}
For the sake of convenience, from now on we use $\Delta$ to denote $\Delta_\mathscr{A}$.
For any $z \in \mathbb{Z}_{\ge 0}$, we denote the set $\{1, \dots, z\}$ by $[z]$.
We subdivide the set of entries in a vector $x \in \mathbb{Z}^{nt}$ into $x$ \textit{bricks} $x^{(1)}, \dots, x^{(n)}$, where each brick is a vector in $\mathbb{Z}^{t}$ and $x^\trans = ((x^{(1)})^\trans, (x^{(2)})^\trans, \dots, (x^{(n)})^\trans)$.
For each $i \in [n], j \in [t]$, we use $x^{(i)}_j$ to denote the entry corresponding to the $j^{th}$ variable in the $i^{th}$ brick of $x$.

For a matrix $M$, we use $M[\cdot, w]$ and $M[w, \cdot]$ to denote the $w^{th}$ column and the $w^{th}$ row of $M$ respectively.
For a matrix $M$, we use $\psum(M,j)$  to denote the partial sum up to column $j$ in $M$: that is, $\psum(M,j) = \sum_{w \in [j]} M[\cdot, w]$.
For a matrix $M$ and an entry $e$, we use $\occ_M(e,j)$ to denote the number of occurrences of $e$ in $M$ up till column $j$.
For a matrix $M$ and a permutation of its columns $\sigma$, we use $M_{\sigma}$ to denote the matrix where $M_\sigma[\cdot, j] = M[\cdot, \sigma(j)]$.

With the definition of bricks, we have that any feasible solution $x$ satisfies
$$
\left(
\begin{array}{cccc}
T^{(1)}    & T^{(2)}    & \cdots & T^{(n)}    \\
1^T    & 0      & \cdots & 0      \\
0      & 1^T    & \cdots & 0      \\
\vdots & \vdots & \ddots & \vdots \\
0      & 0      & \cdots & 1^T    \\
\end{array}
\right)
\left(
\begin{array}{c}
x^{(1)}    \\
x^{(2)}  \\
\vdots\\ \\
x^{(n)}
\end{array}
\right)
=
\left(
\begin{array}{c}
b^{(0)}    \\
b^{(1)}  \\ 
b^{(2)}  \\
\vdots\\ \\
b^{(n)}
\end{array}
\right).
$$
Due to the matrix structure, the problem (\ref{eqn:prob-def}) is equivalent to the following question, which we will focus on from now. 
Let $q = \sum_{i \in [n]} b^{(i)}$.
Does there exist $x \in \mathbb{Z}^{nt}_{\ge 0}$ such that 
\begin{itemize}
    \item~ [all but the top row block] $\sum_{j \in [t]} x_{j}^{(i)}  = b^{(i)}$, for each $i \in [n]$,
    \item~ [top row block] there is a sequence of columns, say $v_1, \dots, v_{q}$, such that $\sum_{k \in [q]} v_k = b^{(0)}$, and column $T^{(i)}[\cdot, j]$ appears $x_{j}^{(i)}$ times in the sequence,
    \item $c^\trans x$ is maximized?
\end{itemize}
If there is no such $x$, then we decide that it is infeasible.
\section{Technical Overview}
\label{sec:overview}
Suppose that the given program \eqref{eqn:prob-def} admits a solution $x$.
From $x$, we obtain a sequence of vectors $v_1, \dots, v_q$ such that
\begin{enumerate}
    \item $q = \sum_{i \in [n]} b^{(i)}$;
    \item each $v_j$ is a column of one of the top blocks $T^{(1)}, \dots, T^{(n)}$;
    \item 
    $
    \sum_{j \in [q]} v_j = b^{(0)};
    $
    \item the column $T^{(i)}[\cdot, k]$ appears $x^{(i)}_k$ many times in the sequence.
\end{enumerate}

Consider the $1 \times q$ matrix $M$ with entries from $[n]$ where $M[j]=i$ if $v_j = T^{(i)}[\cdot, k]$ for some $k \in [t]$, and each $i \in [n]$ is present exactly $b^{(i)}$ times in $M$.
Informally, through $M$ we identify the block from which each vector is obtained.
The first step of our approach is to ``balance'' this one-row matrix $M$.

Suppose that the symbol (or element) $e \in [n]$ appears $m_e$ times in $M$.
Consider a random rearrangement of the matrix $M$.
The expected number of occurrences of $e$ in the first $j$ columns, by the linearity of expectation, is $(j/q) \cdot m_e$.
This is because the probability that a random rearrangement of symbols in $M$ contains $e$ in a particular index is $m_e/q$, and the linearity of expectation is applied over the first $j$ columns.
Let $\occ_{M}(e,j)$ denote the number of occurrences of symbol $e$ between columns $1$ to $j$ in $M$.
The \newdef{imbalance} of a symbol $e$ at a column $j$ in $M$ is defined as 
\[
\imb_{M}(e,j) \coloneq \occ_{M}(e,j) - (j/q) \cdot m_e. \tag{*}\label{def:imbalance}
\]

Thus, the imbalance of a symbol $e$ at a particular column is the difference between the number of occurrences of $e$ and the expected number of occurrences of $e$ until that column in a random rearrangement of the entries.
We show that we can, in time $\bigoh(nq)$, rearrange the symbols such that the matrix has bounded imbalance\footnote{We would like to note that a similar result can be obtained by applying the Steinitz Lemma on the $q*n$ matrix $M'$ where for each $j \in [q]$, $M'[\cdot, j]$ is a vector containing $1$ in position $M[j]$ and $0$ in all other positions.
However, such an application will result in a weaker imbalance upper-bound of $n$ and a slightly more complicated algorithm.}
.

\begin{restatable}[\textbf{Balancing a one-row matrix}]{lemma}{imbalanceboundonerow}
\label{lem:imbalance-bound-one-row}
    Let $M$ be a $1 \times q$ matrix with entries from $[n]$.
    There is a permutation $\sigma$ such that for each symbol $e\in [n]$ and column $j \in [q]$, $-n \le \imb_{M_{\sigma}}(e,j) \le 1$.
    Moreover, the matrix $M_{\sigma}$ can be computed in time $\bigoh(nq)$.
\end{restatable}

For each block $i \in [n]$, we do the following.
Consider the $r \times b^{(i)}$ matrix formed by all the $b^{(i)}$-many vectors in the sequence corresponding to the block $i$, say $V^{(i)}$.
By \Cref{cor:steinitz-lemma-corollary}, there is a permutation of the vectors in $V^{(i)}$, say $\pi^{(i)} \in S_{b^{(i)}}$, such that each coordinate of the $j^{th}$ partial sum is at most $2 r \Delta$ away from the $(j/b^{(i)})^{th}$ fraction of the total sum of vectors in $V^{(i)}$.
We construct a new $r \times q$ matrix $O$ from $M_\sigma$ by replacing the entries with vectors from those blocks.
We place these vectors $V^{(i)}_{\pi^{(i)}}$ (in the same order) in $O$ based on where $i$ occurs in $M_\sigma$: the $w^{th}$ vector in $V^{(i)}_{\pi_{(i)}}$ is placed in the $w^{th}$ occurrence of $i$ in $M_\sigma$.
We show that all partial sums in $O$ are bounded.

\begin{restatable}[\textbf{Bounding the partial sums}]{lemma}{boundpartialsum}
\label{lem:bound-partial-sum}
    For each $j \in [q]$ and $k \in [r]$, we have
    $-n\Delta (n+2r) \le \psum(O,j)[k] - \frac{j}{q}b^{(0)}[k] \le n\Delta (1+2r)$.
\end{restatable}

We construct a weighted directed acyclic graph $G$ based on the bounds we obtained in \Cref{lem:bound-partial-sum}.
The key property of $G$ is that any shortest path between two particular vertices in $G$ corresponds to an optimal solution of the program and vice versa.
If there is no such path, then we assert that there is no solution.
Otherwise, we compute an optimal solution corresponding to the path.
Thus, we have our main result,~\Cref{thm:main}.


\section{Proof of \autoref{thm:main}}

In this section, we prove \Cref{lem:imbalance-bound-one-row} in \Cref{subsec:first-part}, followed by \Cref{lem:bound-partial-sum} in \Cref{subsec:second-part}, and then present the overall algorithm in \Cref{subsec:third-part} which forms a proof of \Cref{thm:main}.

\subsection{Balancing a one-row matrix: Proof of \autoref{lem:imbalance-bound-one-row}}
\label{subsec:first-part}
We restate \Cref{lem:imbalance-bound-one-row} for convenience.

\imbalanceboundonerow*

\begin{algorithm}[h]
    \caption{An algorithm to balance a one-row matrix.}
    \label{alg:construct-balanced-matrix}
    \begin{algorithmic}[1]
    \Require{the number of occurrences of each entry in the $1 \times q$ matrix $M$}
        \State $M_{\sigma} \gets$ an empty $1 \times q$ matrix
        \For{$j$ from $1$ to $q$} \label{line:for-loop-beg}
            \State Let $y \gets \arg\min_{e \in [n]} \imb_{M_{\sigma}}(e,j)$ \label{line:find-most-imbalanced-elt}
            \State Set $M_{\sigma}[j] \gets y$
        \EndFor \label{line:for-loop-end}
        \State \Return $M_{\sigma}$
    \end{algorithmic}
\end{algorithm}

Consider $M_{\sigma}$, the $1\times q$ matrix with entries in $[n]$, constructed by Algorithm \ref{alg:construct-balanced-matrix}.
The algorithm starts with an empty $M_{\sigma}$; then, it iteratively fills the symbols column-wise by finding the symbol with minimum imbalance (note that imbalances can be negative).

Maintaining the imbalances of each of the $n$ distinct symbols and finding a symbol with minimum imbalance (Line \ref{line:find-most-imbalanced-elt}) can be done in time $\bigoh(n)$.
This can be done by storing and updating the imbalances of each of the $n$ symbols.
Overall, Algorithm \ref{alg:construct-balanced-matrix} runs in time $\sum_{j \in [q]} \bigoh(n) = \bigoh(nq)$.

Next, we show the bounds on the symbol imbalances in $M_{\sigma}$.
Recall that a symbol $e \in [n]$ appears $m_e$ times in $M$.
Fix any $j \in [q]$.
Consider the stage of the algorithm where the first $j$ columns of $M_{\sigma}$ are filled.
Now we show the imbalance upper-bound in the lemma.
Observe that the quantity $\imb_{M_{\sigma}}(e,j)$, when viewed as a sequence indexed by $j$, increases only when symbol $e$ is in the entry $M_{\sigma}[j]$.
Assume for the sake of contradiction that $\imb_{M_{\sigma}}(e',j')>1$ for some $e' \in [n]$ and $j' \in [q]$.
Consider the first index $j$ such that $\imb_{M_{\sigma}}(e',j)>1$ and $M_{\sigma}[j]=e'$.
Due to the greedy choice in the construction of $M_{\sigma}$, we have that $\imb_{M_{\sigma}}(e,j)>0$ for each $e \in [n]$.
We then arrive at a contradiction since
\[
j = \sum_{e \in [n]} \occ_{M_{\sigma}}(e,j) = \sum_{e \in [n]} \left( \imb_{M_{\sigma}}(e,j)+\frac{j}{q} m_e \right) > \sum_{e \in [n]} \left( \frac{j}{q} m_e \right) = j.
\]

Next, we show the imbalance lower-bound in the lemma.
A crucial property that we will use to show the bound is that the sum of imbalances at column $j$ is zero.
\begin{claim-inside-lemma}
    \label{clm:imbalance-zero-sum}
    \[
    \sum_{e \in [n]} \imb_{M_{\sigma}}(e,j) = 0.
    \]
\end{claim-inside-lemma}
\begin{proof}
    \[
    \sum_{e \in [n]} \imb_{M_\sigma}(e,j) = \sum_{e \in [n]} \left(\! \occ_{M_{\sigma}}(e,j) - \frac{j}{q} m_e \!\right) = \sum_{e \in [n]} \occ_{M_{\sigma}}(e,j) - \sum_{e \in [n]} \frac{j}{q} m_e = j - \frac{j}{q} q= 0.
    \qedhere
    \]
\end{proof}
Assume for contradiction that $\imb_{M_{\sigma}}(e',j') < -n$ for some $e' \in [n]$ and $j' \in [q]$.
The sum of symbol imbalances at $j'$ is
\[
\sum_{e \in [n]} \imb_{M_{\sigma}}(e,j') = \imb_{M_{\sigma}}(e',j') + \sum_{e \in [n] \setminus \{e'\}} \imb_{M_{\sigma}}(e,j') < -n + \sum_{e \in [n] \setminus \{e'\}} \imb_{M_{\sigma}}(e,j').
\]
Combining the above with \Cref{clm:imbalance-zero-sum}, there exists an element $e \in [n]\setminus \{e'\}$ such that $\imb_{M_{\sigma}}(e,j') > 1$ and thus we have a contradiction.

Combining the lower and upper-bounds, we infer that $-n \le \imb_{M_{\sigma}}(e,j) \le 1$ for each $e \in [n]$ and $j \in [q]$.
What is left to be shown is that the matrix $M_{\sigma}$ is a permutation of the columns of $M$.
For that, it is sufficient to prove that for each $e \in [n]$,
the number of occurrences of element $e$ in $M_{\sigma}$ is $m_e$.
Assume for contradiction that $M_{\sigma}$ is not a permutation of columns of $M$.
We observe that since $\sum_{e \in [n]} \occ_{M_{\sigma}}(e,q) = q = \sum_{e \in [n]} m_e$, there exists a symbol $e' \in [n]$ such that $\occ_{M_{\sigma}}(e',q) > m_{e'}$.
Consider the last index $j$ such that the $j^{th}$ column in $M_{\sigma}$ contains $e'$. 
We have 
\[
\imb_{M_{\sigma}}(e',j) = \occ_{M_{\sigma}}(e',j) - \frac{j}{q}m_{e'} = \occ_{M_{\sigma}}(e',q) - \frac{j}{q}m_{e'} \ge m_{e'} +1 - \frac{j}{q}m_{e'}.
\]
If $j<q$, then $\imb_{M_{\sigma}}(e',j) > 1$ and we have a contradiction to the upper-bound in the lemma.
Otherwise, we have $j=q$. 
Observe that $\imb_{M_{\sigma}}(e',q)=1$.
By Line \ref{line:find-most-imbalanced-elt}, we have that $\imb_{M_{\sigma}}(e,q)\ge 0$ for each $e \in [n] \setminus \{e'\}$.
However, this implies that $\sum_{e \in [n]} \imb_{M_{\sigma}}(e,q) \ge 1$, a contradiction to \Cref{clm:imbalance-zero-sum}.
This completes the proof of the lemma.

\subsection{Bounding the partial sums: Proof of \autoref{lem:bound-partial-sum}}
\label{subsec:second-part}

We restate \Cref{lem:bound-partial-sum} for convenience.

\boundpartialsum*

Before we prove \autoref{lem:bound-partial-sum}, we recall the construction of $O$ and relevant notation that were introduced in \Cref{sec:overview}.
We consider an optimal solution $x$ of (\ref{eqn:prob-def}) and from it obtain a sequence of vectors $v_1, \dots, v_q$ satisfying the conditions given in \Cref{sec:overview}.
Next, using \Cref{alg:construct-balanced-matrix} we compute $M_\sigma$, a $1 \times q$ matrix such that for each symbol $i \in [n]$
\begin{enumerate}
    \item $i$ occurs $b^{(i)}$ times in $M_\sigma$;
    \item for each $j \in [q]$, $-n \le \imb_{M_\sigma}(i,j) \le 1$. (i.e., $-n + (j/q) b^{(i)} \le \occ_{M_\sigma}(i,j) \le 1 + (j/q) b^{(i)}$).
\end{enumerate}

For each block $i \!\in\! [n]$, we do the following. 
Consider $V^{(i)}$, the $r \!\times\! b^{(i)}$ matrix formed by the set of columns in the sequence corresponding to block $i$.
Applying \Cref{cor:steinitz-lemma-corollary} on $V^{(i)}$, we obtain a permutation $\pi_i \in S_{b^{(i)}}$ such that all partial sums of $V^{(i)}_{\pi_i}$ are bounded.
In particular,
\begin{align}
    \lVert 
    \psum (V^{(i)}_{\pi_i}, w) - \frac{w}{b^{(i)}} \cdot \tot(V^{(i)})
    \rVert_\infty
    \le
    2 r\Delta 
    \text{ for all $w \in [b^{(i)}]$},
\end{align}
where $\tot(V^{(i)})$ is the sum of the columns of $V^{(i)}$ (note that $\tot(V^{(i)}) = \tot(V^{(i)}_{\pi_i})$).


\begin{algorithm}[h]
    \caption{An algorithm to construct the matrix $O$.}
    \label{alg:construct-O}
    \begin{algorithmic}[1]
        \State $O \gets$ an empty $r \times q$ matrix
        \State $w_1, \dots, w_n \leftarrow 0$
        \For{$j$ from $1$ to $q$} 
            \State $e \leftarrow M_\sigma[j]$
            \State $w_e \leftarrow w_e + 1$
            \State $O[\cdot, j] \leftarrow V^{(i)}_{\pi_i}[\cdot, w_e]$
        \EndFor 
        \State \Return $O$
    \end{algorithmic}
\end{algorithm}

Next, we construct $O$, an $r \times q$ matrix as described in \Cref{alg:construct-O}.
Essentially, we place the columns of $V^{(i)}_{\pi_i}$ on the positions where $i$ appears in $M_\sigma$.
Now we are ready to show that this matrix has bounded partial sums.

For any $j \in [q]$ and $k \in [r]$, we bound the partial sum of $O$ at index $k$ of column $j$ as

\begin{align*} 
\psum(O,j)[k] &= \sum_{i \in [n]} 
\psum(V^{(i)}_{\pi_i}, \occ_{M_\sigma}(i,j))[k]
\le \sum_{i \in [n]} \left( \frac{\mathrm{occ}(i,j)}{b^{(i)}} \tot(V^{(i)}_{\pi_i})[k] + 2r \Delta \right) &\\
&\le \sum_{i \in [n]} \left( \frac{1 + \frac{j}{q}b^{(i)}}{b^{(i)}} \tot(V^{(i)})[k] + 2r \Delta \right) &\\
&= \sum_{i\in[n]} \left( \frac{1}{b^{(i)}}\tot(V^{(i)})[k] + \frac{j}{q}\tot(V^{(i)})[k] + 2r \Delta \right) &\\
&\le \sum_{i\in[n]} \left( \frac{1}{b^{(i)}}\Delta b^{(i)} + \frac{j}{q}\tot(V^{(i)})[k] + 2r \Delta \right)
= \frac{j}{q}b^{(0)}[k] + \sum_{i\in[n]} \left( \Delta + 2r \Delta \right) &\\
&= \frac{j}{q}b^{(0)}[k] + n\Delta(1+2r).
\end{align*}


Similarly, we have the following lower bound.
\begin{align*} 
\psum(O,j)[k] &= \sum_{i \in [n]} 
\psum(V^{(i)}_{\pi_i}, \occ_{M_\sigma}(i,j))[k]
\ge \sum_{i \in [n]} \left( \frac{\mathrm{occ}(i,j)}{b^{(i)}} \tot(V^{(i)}_{\pi_i})[k] - 2r \Delta \right) &\\
&\ge \sum_{i \in [n]} \left( \frac{-n + \frac{j}{q}b^{(i)}}{b^{(i)}} \tot(V^{(i)})[k] - 2r \Delta \right) &\\
&= \sum_{i\in[n]} \left( \frac{-n}{b^{(i)}}\tot(V^{(i)})[k] + \frac{j}{q}\tot(V^{(i)})[k] - 2r \Delta \right) &\\
&\ge \sum_{i\in[n]} \!\left(\! \frac{-n}{b^{(i)}}\Delta b^{(i)} + \frac{j}{q}\tot(V^{(i)})[k] - 2r \Delta \right)\! 
= \frac{j}{q}b^{(0)}[k] \!+\! \sum_{i\in[n]} \!\left( -n\Delta - 2r \Delta \right) &\\
&= \frac{j}{q}b^{(0)}[k] - n\Delta(n+2r).
\end{align*}

\subsection{The algorithm}
\label{subsec:third-part}
\begin{algorithm}[h]
    \caption{An algorithm to solve (\ref{eqn:prob-def}).}
    \label{alg:overall-algorithm}
    \begin{algorithmic}[1]
        \State Compute $M_{\sigma}$ using \Cref{alg:construct-balanced-matrix}.
        \State Construct the graph $G = (V,A)$ as given in \Cref{const:graph}.
        \State Compute a shortest path $P$ between $h_{0,0}$ and $h_{q,b^{(0)}}$.
        \State Return an optimal solution corresponding to $P$.
    \end{algorithmic}
\end{algorithm}

We now describe a direct algorithm to solve (\ref{eqn:prob-def}). 
From now on, we suppose that the given program admits an optimal solution $x^*$.
Otherwise, the algorithm asserts that there is no solution for the program and we are correct.
Firstly, using \Cref{alg:construct-balanced-matrix}, we compute $M_\sigma$, a $1 \times q$ matrix that satisfies the conditions of \Cref{lem:imbalance-bound-one-row}.
Note that the existence of $x^*$ alone is sufficient to compute $M_\sigma$ since \Cref{alg:construct-balanced-matrix} only requires the total number of occurrences of each symbol (which is $b^{(i)}$ for each $i \in [n]$).

The existence of $x^*$ now implies that there is a matrix $O$ which satisfies the conditions of \Cref{lem:bound-partial-sum}.
To find such an $x^*$, we find the existence of a matrix satisfying \Cref{lem:bound-partial-sum}, by finding shortest paths in certain graphs - in a way similar to \cite{DBLP:journals/talg/EisenbrandW20}.
We construct a weighted directed acyclic graph $G = (V,A)$ as follows.

\begin{construction}
\label{const:graph}
~
\begin{enumerate}
    \item (Vertices) For each $j \in [n]$ and each integer vector $v \in \mathbb{Z}^r$ such that $$(j/q) b^{0}[k] - n\Delta(n+2r) \le v[k] \le (j/q) b^{0}[k] + n\Delta(1+2r)$$ for each $k \in [r]$, we add a vertex $h_{j,v}$ to $V$.
    Lastly, we add the vertex $h_{0,0}$ to $V$.
    \item (Edges) For each $j \in [q]$, we do the following.
    Suppose that $M_\sigma[j] = i$.
    We add the arc $(h_{v_1, j-1}, h_{v_2,j})$ to $A$ if $v_2-v_1$ is the column $T^{(i)}[\cdot, k]$ for some $k \in [t]$, and the weight of the arc is set as $c^{(i)}_k$.
\end{enumerate}
\end{construction}
In the following claim, we bound the sizes of $|V|$ and $|A|$. 

\begin{claim}
    $|V| \le 1 + q \cdot (n\Delta(n+1+4r))^r$ and $|A| = \bigoh(nt |V|)$.
    Moreover, we can construct $G$ in time $\bigoh(qnt \cdot (n\Delta(n+1+4r))^r)$.
\end{claim}
\begin{claimproof}
    For each $j \in [q]$, the number of integer vectors $v$ such that $(j/q) b^{0}[k] - n\Delta(n+2r) \le v[k] \le (j/q) b^{0}[k] + n\Delta(1+2r)$ for each $k \in [r]$ is at most $(n\Delta(n+1+4r))^r$.
    This is because for a fixed $k \in [r]$, the number of possible values that $v[k]$ can take is $n\Delta(n+1+4r)$.
    Summing over all $j \in [q]$ and then counting the vertex $h_{0,0}$, we have the claim.
    Note that we can construct $V$ in time proportional to its size.

    To construct $A$, we can iterate over all vertices $h_{j,v}$, and in time $nt$ compute all arcs incident on $h_{j,v}$.
    Thus, we have $|A| = \bigoh(nt |V|)$, and we can construct the graph $G$ in time $\bigoh(|V| + nt |V|) = \bigoh(qnt \cdot (n\Delta(n+1+4r))^r)$.
\end{claimproof}


By the definition of $G$ and the existence of $x^*$, we have that there is a path of cost $c^\trans x^*$ in $G$ between vertices $h_{0,0}$ and $h_{b^{(0)}, q}$ with cost $c^\trans x^*$ and this is the shortest such path.
We can compute such a path by using $BFS$ in time $\bigoh(|A|) = \bigoh(qnt \cdot (n\Delta(n+1+4r))^r)$.
To compute an optimal solution for (\ref{eqn:prob-def}) corresponding to $P$, we simply trace the path and set the variables according to the vertices that appear on the path.

\section{Dealing with inequalities in constraints}\label{ss:dealing-with-inequalities}
In this section, we generalize our main result by showing that programs which contain ``certain'' inequalities in the constraints can be solved in the same time that we take for \eqref{eqn:prob-def} where we have $\mathscr{A}x = b$.
This is towards the ease of expressing problems involving inequalities as $n$-fold integer programs which we will see later in \Cref{sec:Applications}. 


We call the upper rows $(T^{(1)} ~ T^{(2)} ~ \cdots ~ T^{(n)})x = b^{(0)}$ as \textit{globally uniform constraints}, and the lower rows $1^\trans x^{(i)} = b^{(i)}$ as \textit{locally uniform constraints}.
Consider the generalization of \eqref{eqn:prob-def} \begin{enumerate}
    \item\label{item:globally-uniform-constraints} where the globally uniform constraints may contain either of  $\{ \le, = , \ge \}$;
    \item\label{item:locally-uniform-constraints} where the locally uniform constraints may contain either of $\{ \le, = \}$.
\end{enumerate}
Let $A = (1, \dots, 1) \in \mathbb{Z}^{1 \times t}$, $T^{(1)}, \dots, T^{(n)} \in \mathbb{Z}^{r \times t}$, $b \in \mathbb{Z}^{r+n}$, and $c \in \mathbb{Z}^{nt \times 1}$. 
Formally, we consider the following generalization.
\begin{align}
\label{eqn:prob-def-with-ineq}
&\min\left\{c^\trans x\, \mid \mathscr{A} x ~\diamond~ b\,, x\in\mathbb{Z}_{\ge 0}^{nt}\right\} \tag{$\mathbf{P_2}$}\\
&~~\mbox{where }~\mathscr{A}:=
\left(
\begin{array}{cccc}
T^{(1)}    & T^{(2)}    & \cdots & T^{(n)}    \\
A    & 0      & \cdots & 0      \\
0      & A    & \cdots & 0      \\
\vdots & \vdots & \ddots & \vdots \\
0      & 0      & \cdots & A    \\
\end{array}\notag
\right)\\
&~~\text{and $\diamond$ is a sequence of inequalities satisfying conditions \ref{item:globally-uniform-constraints} and \ref{item:locally-uniform-constraints}}.\notag
\end{align}
We emphasize that the generalization does not allow locally uniform constraints to contain ``$\ge$''.
We would like to note that our approach at a high level is almost the same as that of \cite{DBLP:journals/mp/KnopKM20}, and the main difference lies in the fact that (unlike as in their model,) here the variables in the programs we consider are non-negative and have no explicit lower or upper bounds.
The proof strategy is to construct an appropriate program of the form \eqref{eqn:prob-def} by adding new (slack) variables.
Let $\psi \coloneq 2 + 2\lVert c \rVert_\infty \cdot \max_{i \in [n]} \{ b^{(i)} \}$. 
Observe that $\psi/2$ is strictly larger than the maximum value the objective function in \eqref{eqn:prob-def-with-ineq} can take.

First, we deal with the locally uniform constraints in \eqref{eqn:prob-def-with-ineq}.
We add $n$ variables:
for each $i \in [n]$, we add $x^{(i)}_{t+1}$ and replace $T^{(i)}$ by $(T^{(i)} ~ \mathbf{0})$ where $\mathbf{0}$ is the $r \times 1$ matrix (i.e column) containing zeros.
If the $i^{th}$ constraint has a ``$\le$'', then we set $c^{(i)}_{t+1} = 0$.
Otherwise it contains an ``$=$'', and we set $c^{(i)}_{t+1} = \psi$.

Next, we deal with the globally uniform constraints in \eqref{eqn:prob-def-with-ineq}.
Without loss of generality, we assume that each constraint either contains a ``$\le$'' or an ``$=$''.
This is a safe assumption because if there is a constraint containing a ``$\ge$'', then we can negate that constraint (which flips the inequality in that constraint to ``$\le$'') and obtain a program that is still of the form we consider.
We replace each $(T^{(i)} ~ \mathbf{0})$ by $( T^{(i)} ~ \mathbf{0} ~ I_r)$, where $I_r$ is the $r \times r$ identity matrix.
We also introduce a new $(n+1)^{st}$ block $(Z_r ~ \mathbf{0} ~ I_r)$ where $Z_r$ is the $r \times t$ matrix consisting of zeros, and $\mathbf{0}$ is the $r \times 1$ matrix consisting of zeros.
The coefficients of the new variables in the first $n$ blocks are set to $\psi$.
That is, for each $i \in [n]$ and $j \in [t+2, t+r+1]$, we set $c^{(i)}_j = \psi$.
For the $(n+1)^{st}$ block, we set the coefficients as follows.
For each $j \in [t+r+1]$, we set $c^{(n+1)}_j = 0$.
Then, we set $b^{(n+1)} = 2r\Delta \cdot \max_{i \in [n]} \{ b^{(i)} \}$, where $\Delta$ is the largest entry in the constraint matrix $\mathscr{A}$ we started with.
This concludes the construction of the program of the form \eqref{eqn:prob-def}.

Consider any feasible solution $y$ to the initial program \eqref{eqn:prob-def-with-ineq}.
We define a solution $x$ to the constructed program of the form \eqref{eqn:prob-def} of the same cost by setting the newly added variables appropriately.
Firstly, we set $x^{(i)}_j = y^{(i)}_j$ for each $i \in [n]$ and $j \in [t]$.
Recall that we assume w.l.o.g. that each globally uniform constraint contains either a ``$\le$'' or an ``$=$''. 
For each $k \in [r]$, we deal with the $k^{th}$ globally uniform constraint in \eqref{eqn:prob-def-with-ineq} as follows.
We set $x^{(n+1)}_{t+1+k} = b^{(0)}[k] - \mathscr{A}[k, \cdot] y$.
Observe that $x^{(n+1)}_{t+1+k}$ is precisely the slack in the $k^{th}$ globally uniform constraint corresponding to solution $y$, and it is a number that is at most $2\Delta \cdot b^{(k)}$.
For each $i \in [n]$, we deal with the $i^{th}$ locally uniform constraint in \eqref{eqn:prob-def-with-ineq} as follows.
We set $x^{(i)}_{t+1} = b^{(i)} - (1 \dots 1)^\trans y^{(i)}$.
Finally, we set $x^{(n+1)}_{t+1} = b^{(n+1)} - \sum_{k \in [r]} x^{(n+1)}_{t+1+k}$.
Similar to before, observe that $x^{(i)}_{t+1}$ is precisely the slack in the $i^{th}$ locally uniform constraint corresponding to solution $y$, and it is a number that is at most $b^{(i)}$.
All other variables whose values have not yet been set in $x$, we set them to $0$.

We claim that $x$ is a solution to the constructed program \eqref{eqn:prob-def} of the same cost.
Firstly, it is a feasible solution by construction.
Secondly, the cost is the same because any newly added variable that is set to a non-zero value corresponds to a ``$\le$'' constraint, and its coefficient is $0$ in \eqref{eqn:prob-def}. 
For the other direction,
suppose that we are given a solution $x$ to \eqref{eqn:prob-def}.
If the cost of $x$ is at least $\psi/2$ (which is strictly larger than the value of any feasible solution in \eqref{eqn:prob-def-with-ineq}), then we can assert that \eqref{eqn:prob-def-with-ineq} does not admit any solution. 
Otherwise, we obtain a solution to \eqref{eqn:prob-def-with-ineq} of the same cost by discarding the newly added variables.

Now that we have established how to compute the solution of one program given that of the other, we move on to assess the time to construct \eqref{eqn:prob-def} and the change in parameters.
Each locally uniform constraint can be dealt with separately in time $\bigoh(nt)$ by adding a column corresponding to it and updating $c$ appropriately.
All globally uniform constraints can be dealt with together in time $\bigoh(n \cdot rnt)$ by adding $r$ columns to each block, adding a new block, and then updating $b$ appropriately.
In total, we can construct \eqref{eqn:prob-def} in time $\bigoh(n^2 rt)$. 
Thus, given \eqref{eqn:prob-def-with-ineq}, we can construct \eqref{eqn:prob-def} in time $\bigoh(n^2rt)$.

Next, we assess the change in parameters: the number of blocks is $n+1$, the number of columns in a block is $t+r+1$, the number of globally uniform constraints is $r$, the parameter $q$ (which is the sum of the right hand sides of the locally uniform constraints) becomes $q + 2r\Delta \max_{i \in [n]} \{ b^{(i)} \}$, and the largest entry $\Delta$ remains the same since the newly added entries to the matrix are either $0$ or $1$.
Applying \Cref{thm:main}, we can solve \eqref{eqn:prob-def} in time 
\begin{align}
    \bigoh( (q + 2r\Delta \max_{i \in [n]} \{ b^{(i)} \} ) (n+1) (t+r+1) \cdot ((n+1)\Delta(n+1+1+4r))^r) \\
    \le \bigoh( qr\Delta (n+1) (t+r+1) \cdot ((n+1)\Delta(n+2+4r))^r)\\
    \le \bigoh( qrt \cdot ((n+1)\Delta(n+2+4r))^{r+1}).
\end{align}
Summing with the time to construct \eqref{eqn:prob-def}, we obtain the total time to solve \eqref{eqn:prob-def-with-ineq}.


\begin{corollary}
    \label{cor:main-result}
    The ILP \eqref{eqn:prob-def-with-ineq} can be solved in time $\bigoh(qrt \cdot ((n+1)\Delta(n+2+4r))^{r+1})$.
\end{corollary}

\section{Applications}\label{sec:Applications}
In this section, we apply our main result on three problems: {\sc Lobbying}, \textsc{Binary} \closeststring, and \equitablecoloring.

\subsection{Lobbying in multiple referenda}
\label{subsec:application1}
Given the binary approvals (equivalently, yes or no answers) of $n$ voters on $m$ issues, the question of \lobbying is whether the lobby can choose $k$ voters to be ``influenced'' so that each issue gets a majority of approvals.
By influence a voter, we mean gaining complete control over the voter, and getting her approval on all issues.
Christian et al. \cite{Christian2007} introduced \lobbying and modeled it as the following binary matrix modification problem.

\defproblem{\lobbying}
{A matrix $A \in \{0,1\}^{w \times m}$ and an integer $k \ge 0$}
{Can one choose $k$ rows and flip all 0s to 1s in these rows so that in the resulting matrix every column has more 1s than 0s?}

First, we recall how Bredereck et al. \cite{DBLP:journals/jair/BredereckCHKNSW14} express \lobbying as an integer program.
Let $\ell$ denote the number of distinct types of rows in $M$: two rows are said to be of the same type if they are identical.
Observe that $\ell \le 2^m$.
Let $c(r_1), \dots, c(r_\ell)$ denote the number of rows of each type.
For each $i \in [\ell]$ and $j \in [m]$, let $B_j(r_i) = 1$ if the $j^{th}$ column of row type $r_i$ has value $0$, and let $B_j(r_i) = 0$ otherwise.
For each $i \in [\ell]$, let $b_i$ be an integer variable that indicates the number of times one has to modify a row of type $r_i$.
For each $j \in [m]$, let $g_j$ denote the number of missing 1s to make column $j$ have more 1s than 0s.
Thus, we have the following two constraints: 
(1) $0 \le b_i \le c(r_i)$ for each $i \in [\ell]$; 
(2) $g_j \le \sum_{i \in [\ell]} b_i \cdot B_j(r_i)$ for each $j \in [m]$.
Overall, we have the following program which is of the form \eqref{eqn:prob-def-with-ineq}.
The objective is to minimize $\sum_{i \in [\ell]} b_i$ subject to
\[
\left(
\begin{array}{cccc}
-B_1(r_1)    & -B_1(r_2)    & \cdots & -B_1(r_\ell)    \\
-B_2(r_1)    & -B_2(r_2)    & \cdots & -B_2(r_\ell)    \\
\vdots    & \vdots    & \ddots & \vdots    \\
-B_m(r_1)    & -B_m(r_2)    & \cdots & -B_m(r_\ell)    \\
1         & 0          & \cdots & 0      \\
0           & 1    & \cdots & 0      \\
\vdots      & \vdots     & \ddots & \vdots \\
0           & 0          & \cdots & 1   \\
\end{array}
\right)
\left(
\begin{array}{c}
b_1    \\
b_2  \\
\vdots\\ \\
b_\ell
\end{array}
\right)
\le
\left(
\begin{array}{c}
-g_1    \\
-g_2    \\
\vdots    \\
-g_m    \\
c(r_1)  \\ 
c(r_2)  \\
\vdots\\ \\
c(r_\ell)
\end{array}
\right).
\]

Observe that the optimal solution is at most $k$ if and only if there are $k$ rows that can be chosen to satisfy the objective in question.
The parameters corresponding to the above program are $r=m$, $t=1$, $n=\ell \le 2^m$, $\Delta=1$, and $q = \sum_{k \in [m]} g_k + \sum_{i \in [\ell]} c(r_\ell) \le wm + w = (m+1)w$.
Given the input, we can construct the above program in time $\bigoh(2^m w)$.
Applying \Cref{cor:main-result}, we have the following.

\begin{theorem}
    \lobbying can be solved in time $2^{\bigoh(m^2)}\cdot w^{\bigoh(1)}$.
\end{theorem}
The best known dependency on $m$ prior to this work, established by Knop et al. \cite{DBLP:journals/mp/KnopKM20}, is $m^{\bigoh(m^2)}$.
Next, we discuss the consequence of existing algorithms on the above program.
Using the algorithms for standard ILPs, including \cite{DBLP:journals/mor/Kannan87,DBLP:journals/mor/JansenR23}, results in algorithms that are double exponential with respect to $m$.
Using the algorithms for \nfILP, including \cite{DBLP:journals/siamdm/JansenLR20,DBLP:conf/soda/CslovjecsekEHRW21}, results in algorithms that run in time $m^{\bigoh(m^2)}\log w + \bigoh(2^m \cdot w)$ (the $\bigoh(2^m \cdot w)$ term is for the construction of the program).
Thus, we establish an improvement in running time in terms of the dependence on $m$.


\subsection{Stringology}
We consider $\dms$, a general problem defined by Knop et al. in \cite{DBLP:journals/mp/KnopKM20} that captures many previously studied string problems including \textsc{Closest String}, \textsc{Farthest String}, \textsc{$d$-Mismatch}, \textsc{Distinguishing String Selection}, \textsc{Optimal Consensus}, and \textsc{Closest to Most Strings}.
See \cite{DBLP:journals/mp/KnopKM20} for a clear description of how it captures the other problems.

\defproblem{\dms}
{$k$ strings $s_1, \dots, s_k$, each of length $L$ from an alphabet $\Sigma \cup \{\star\}$ (where $\star$ denotes a wildcard: a special character that matches with all characters in $\Sigma$), distance lower-bounds $d_1, \dots, d_k \in \mathbb{N}$, and distance upper-bounds $D_1, \dots, D_k \in \mathbb{N}$, distance function $\delta \colon \Sigma^* \times \Sigma^* \rightarrow \mathbb{N}$, and a binary parameter $\beta \in \{0,1\}$.}
{Find a string $y \in \Sigma^L$ such that 
\begin{itemize}
    \item $y$ minimizes $\beta \cdot \left( \sum_{h=1}^{k} \delta(y,s_h) \right)$;
    \item for each $s_h$, the distance (based on $\delta$) between $y$ and $s_h$, denoted by $\delta(y,s_h)$, is at least $d_h$ and at most $D_h$.
\end{itemize}}

A restriction of $\delta$ that makes the problem expressible by $n$-fold integer programs is when $\delta$ is a \emph{character-wise wildcard-compatible}: for any two strings $x,y \in (\Sigma \cup \{\star\})^L$, $\delta(x,y) = \sum_{j=1}^{L} \delta(x[j],y[j])$ and $\delta(c, \star)=0$ for all $c \in \Sigma$.
From now on, we consider only this restriction.

We begin by viewing the input as a $k\times L$ character matrix $S$ where the $i^{th}$ row contains the string $s_i$.
We say that two columns in $S$ are of the same type if they are identical (i.e., both in terms of the number of occurrences of each symbol and the order of the occurrences). 
Since $A$ is a $k \times L$ matrix with entries from $\Sigma \cup \{\star\}$, there are at most $w=(|\Sigma| + 1)^k$ column-types in $S$. 
W.l.o.g., we assume that each column-type is denoted by an element in $[w]$.
A solution $z \in \Sigma^L$ can be mapped to a $(k+1) \times L$ matrix where the first $k$ rows are the $k$ input strings and the $(k+1)^{st}$ row is $z$.
We construct an integer program that detects the existence of such a matrix.

For each $i \in [w]$ and $c \in \Sigma$, the variable $x^{(i)}_c$ indicates the number of columns in the solution matrix which contain $i$ followed by $c$.
For each $i \in [w]$, let $\mathbf{e_i}$ denote the $k \times 1$ string corresponding to column-type $i$.
The globally uniform constraints are the following.
For each $j \in [k]$, we have 
\[
\sum_{c \in \Sigma} \sum_{i \in [w]} x^{(i)}_{c} ~\delta(c, \mathbf{e_i}[j]) \ge  d_j;
\quad \quad 
\sum_{c \in \Sigma} \sum_{i \in [w]} x^{(i)}_{c} ~\delta(c, \mathbf{e_i}[j]) \le  D_j.
\]
For each $i \in [w]$, let $o_i$ denote the number of columns of type $i$; the $i^{th}$ locally uniform constraint constraint is 
\[
\sum_{c \in \Sigma} x^{(i)}_c = o_i.
\]
The objective function is $\beta \cdot \left( \sum_{h=1}^{k} \delta(y,s_h) \right) = \beta \cdot \left( \sum_{h \in [k]}\sum_{i \in [w]} \sum_{c \in \Sigma} x^{(i)}_c \delta(\mathbf{e}_i[h], c)\right)$.
Note that it is a linear function on the variables since $\beta \in \{0,1\}$, and thus we can set $c$ appropriately.
The parameters corresponding to the program are $n = (|\Sigma|+1)^k$, $r = 2k$, $t = |\Sigma|$, $q = \sum_{i \in [w]} o_i = L$, and $\Delta = \max_{c_1, c_2} \delta(c_1, c_2)$.
Given the input, we can construct the above program in time $\bigoh((|\Sigma|+1)^k kL)$.
Applying \Cref{cor:main-result}, we have the following.

\begin{theorem}
\label{thm:Multistring time complexity}
    \dms can be solved in time 
    $(\Delta k)^{\bigoh(k)}(|\Sigma|)^{\bigoh(k^2)} \cdot L^{\bigoh(1)}$,
    where $\Delta = \max_{c_1,c_2} \delta(c_1,c_2)$, and $\delta$ is a character-wise wildcard-compatible function.
    
    
\end{theorem}

\begin{sloppypar}
Knop et al. \cite{DBLP:journals/mp/KnopKM20} showed that \dms can be solved in time $((|\Sigma|+1)^{\bigoh(k^2)} (\Delta k)^{\bigoh(k^2)}\log (L) + \bigoh((|\Sigma|+1)^k kL)$.
Subsequent algorithms for \nfILP applied on the program they consider, including \cite{DBLP:journals/siamdm/JansenLR20,DBLP:conf/soda/CslovjecsekEHRW21}, result in algorithms that run in time $(k\Delta)^{\bigoh(k^2)}(|\Sigma|+1)^{\bigoh(k)} + \bigoh((|\Sigma|+1)^k kL)$.
Next, we discuss the application of existing algorithms in the above program.
Using the algorithms for standard ILPs, including \cite{DBLP:journals/mor/Kannan87,DBLP:journals/mor/JansenR23}, results in algorithms that are double exponential in $k$.
Using the algorithm by Cslovjecsek et al. \cite{DBLP:conf/soda/CslovjecsekEHRW21} results in an algorithm that runs in time $(k\Delta)^{\bigoh(k^2)}(|\Sigma|+1)^{\bigoh(k)} + \bigoh((|\Sigma|+1)^k kL)$.
\end{sloppypar}

The problem \closeststring is the special case of \dms where $\beta=0$, the distance function $\delta$ is the Hamming distance, and $d_i = 0$ and $D_i=d$ for each $i \in [k]$.
Applying \Cref{thm:Multistring time complexity}, we obtain an algorithm for \closeststring that runs in time $(k)^{\bigoh(k^2)} \cdot L^2$.
In terms of the dependence on $k$, this matches the current best algorithms \cite{DBLP:conf/focs/EisenbrandRW24,DBLP:journals/mp/KnopKM20}.

When $\Sigma=\{0,1\}$ in the \closeststring problem, \Cref{thm:Multistring time complexity} gives an algorithm that runs in time $2^{\bigoh(k^2)} \cdot L^2$.
We note that this is an improvement over applying the existing algorithms for \nfILP, including \cite{DBLP:journals/siamdm/JansenLR20,DBLP:conf/soda/CslovjecsekEHRW21} which have a $k^{\bigoh(k^2)}$ term in the running time.
To the best of our knowledge, we are the first to exhibit an algorithm with $2^{\bigoh(k^2)}$ dependence on $k$.
We note that this matches the lower bound hypothesized by Rohwedder and Wegrzycki \cite{DBLP:conf/innovations/RohwedderW25} for the problem.
Thus, we have the following.
\begin{corollary}
    \closeststring can be solved in time $(k)^{\bigoh(k^2)} \cdot L^{\bigoh(1)}$. Moreover, \textsc{Binary} \closeststring can be solved in time $2^{\bigoh(k^2)}\cdot L^{\bigoh(1)}$.
\end{corollary}


\subsection{Equitable coloring parameterized by vertex cover}

We consider the \equitablecoloring problem parameterized by the vertex cover number $k$.
We note that Gomes et al. \cite{DBLP:journals/algorithmica/GomesGS23} show that the problem can be solved in time $2^{\bigoh(k \log k)} \cdot |V(G)|^{\bigoh(1)}$ using a flow-based approach. 
Though we don't match their running time, we still present an ILP to exhibit the applicability of our algorithm.

Given a graph $G=(V,E)$, a \textit{vertex coloring} of $G$ is a function $c: V(G) \mapsto \mathbb{N}$ if for any $(u,v) \in E(G)$ we have $c(u) \ne c(v)$.
A vertex $v$ is said to be \textit{colored} $i$ if $c(v)=i$.
For each $i$, the set of vertices colored $i$ form the $i^{th}$ color class $V_i$.
We study the following problem.

\defproblem{\equitablecoloring}
{A graph $G=(V,E)$ and a positive integer $h$}
{Is there a vertex coloring $c$ of $G$ using at most $h$ colors such that the sizes of any two color classes differ by at most $1$.}

A \textit{vertex cover} in $G$ is a subset of vertices whose deletion results in an edgeless graph (also called an \textit{independent set}).
The \textit{vertex cover number} of $G$ is the size of a smallest vertex cover in $G$.
Fiala et al. \cite{DBLP:journals/tcs/FialaGK11} show that \equitablecoloring is \FPT when with respect to the vertex cover number $k$.
They study the two cases of $h\le k$ and $h \ge k$ separately. 
For the sake of presentation, we only cover the first case, and we note that the second case is dealt with in an almost identical manner and the claimed running time holds.

We assume that we are given a vertex cover $W$ of size $k$: note that such a $W$ can be computed in time $2^{\bigoh(k)}$ \cite{DBLP:books/sp/CyganFKLMPPS15}.
Let $\{I_1, \dots, I_s\}$ denote the partition of the independent set $V(G) \setminus W$ according to their neighborhoods. 
Observe that $s \le 2^k$.

Let $w = \lfloor \frac{|V(G)|}{h} \rfloor$, $a = |V(G)|-hw$, and $b = h-a$.
For each proper coloring $(V_1, \dots, V_h)$ of $W$, they construct a system of linear inequalities with $sh$ variables $x^{(i)}_{j}$, $i \in [s]$ and $j \in [h]$, where $x^{(i)}_j$ indicates the number of vertices of color $j$ in the set $I_i$:

\begin{enumerate}
    \item $x^{(i)}_{j} \ge 0$;
    \item\label{item:ec-global1} $x^{(i)}_j = 0$, if color $j$ is used in $N(I_i)$. Note that the non-negativity of the variables implies that this can equivalently be expressed as a single constraint $\sum_{i,j} x^{(i)}_{j} = 0$ where the summation is over $i\in [s], j \in [h]$ such that color $j$ is used in $N(I_i)$;
    \item\label{item:ec-global2} $x^{(1)}_{j} + \dots + x^{(s)}_{j} = t+1-|W \cap V_j|$ if $j \in \{1. \dots, a\}$;
    \item\label{item:ec-global3} $x^{(1)}_{j} + \dots + x^{(s)}_{j} = t-|W \cap V_j|$ if $j \in \{a, \dots, h\}$;
    \item\label{item:ec-local} $x^{(i)}_{1} + \dots + x^{(i)}_{h} = |I_i|$ for each $i \in [s]$.
\end{enumerate}
We claim that the above integer program is of the form \eqref{eqn:prob-def-with-ineq}.
The $sh$ variables are partitioned into $s$ blocks of size $h$.
The constraints written in \Cref{item:ec-global1,item:ec-global2,item:ec-global3} form the $h+1$ global constraints.
The constraints written in \Cref{item:ec-local} form the local constraints.
The parameters corresponding to the above program are $r=k+1$, $t = h \le |V(G)|$, $n = s \le 2^k$, $\Delta=1$, and $q = \sum_{i \in [s]} |I_i| = |V(G)|-k$.
Recall that for each $k$-coloring of $W$ an integer program of the above kind is created.
This can be done in time $2^{\bigoh(k)} \cdot |V(G)|^{\bigoh(1)}$.
Thus, the overall running time is given by summing over all $k^k$ colorings the time to create and solve the corresponding program.
Applying \Cref{cor:main-result}, we have the following.
\begin{theorem}
    \equitablecoloring can be solved in time $2^{\bigoh(k^2)} \cdot |V(G)|^{\bigoh(1)}$.
\end{theorem}

We note that using existing algorithms for \nfILP, including \cite{DBLP:journals/siamdm/JansenLR20,DBLP:conf/soda/CslovjecsekEHRW21}, results in a $k^{\bigoh(k^2)}$ dependence on $k$ while we only have a $2^{\bigoh(k^2)}$ dependence.




\bibliography{bibliography}

\end{document}